\documentclass[10pt, numberedappendix]{iopart}
\usepackage{graphicx}
\usepackage{amsfonts}
\usepackage{amssymb}
\usepackage{iopams}
\usepackage{threeparttable}
\begin{document}
\title[High-order Ho multipoles in HoB$_2$C$_2$]{High-order Ho multipoles in HoB$_2$C$_2$ observed with soft resonant x-ray diffraction.}
\date{\today}
\author{A~J~Princep$^{1,2}$, A~M~Mulders$^{1,2,3}$, E~Schierle$^4$, E~Weschke$^4$,J~Hester$^3$,W~D~Hutchison$^1$ Y~Tanaka$^5$,N~Terada$^6$, Y~Narumi$^7$, T~Nakamura$^8$}
\address{$^1$ School of PEMS, UNSW, Canberra, ACT, 2600, Australia}
\address{$^2$ Department of Imaging and Applied Physics, Curtin University, Perth, WA, 6845, Australia}
\address{$^3$ Bragg Institute, ANSTO, Lucas Heights, NSW, 2234, Australia}
\address{$^4$ Helmholtz Zentrum Berlin Mat \& Energie GmbH, D-12489 Berlin, Germany}
\address{$^5$ RIKEN SPring-8 Center, Harima Institute, Sayo, Hyogo 679-5148, Japan}
\address{$^6$ National Institute for Materials Science, Sengen 1-2-1, Tsukuba, Ibaraki 305-0044, Japan }
\address{$^7$ Institute for Materials Research, Tohoku University,  Katahira 2-1-1, Aoba, Sendai 980-8577, Japan}
\address{$^8$ SPring-8/JASRI, Sayo, Hyogo 679-5198, Japan}
\ead{A.Mulders@ADFA.edu.au}
\begin{abstract}
Soft resonant x-ray Bragg diffraction (SRXD) at the Ho M$_{4,5}$ edges has been used to study Ho $4f$ multipoles in the combined magnetic and orbitally ordered phase of HoB$_2$C$_2$.  A full description of the energy dependence for both $\sigma$ and $\pi$ incident x-rays at two different azimuthal angles, as well as the ratio $I_\sigma/I_\pi$ as a function of azimuthal angle for a selection of energies, allows a determination of the higher order multipole moments of rank 1 (dipole) to 6 (hexacontatetrapole). The Ho $4f$ multipole moments have been estimated, indicating a dominant hexadecapole (rank 4) order with an almost negligible influence from either the dipole or the octupole magnetic terms. The analysis incorporates both the intra-atomic magnetic and quadrupolar interactions between the $3d$ core and $4f$ valence shells as well as the interference of contributions to the scattering that behave differently under time reversal. Comparison of SRXD, neutron diffraction and non resonant x-ray diffraction shows that the magnetic and quadrupolar order parameter are distinct. The $(0 0 \frac{1}{2})$ component of the magnetic order exhibits a Brillouin type increase below the orbital ordering temperature T$_Q$, while the quadrupolar order increases more sharply. We conclude the quadrupolar interaction is strong, but quadrupolar order only occurs when the magnetic order gives rise to a quasi doublet ground state, which results in a lock-in of the orbitals at T$_Q$.

\end{abstract}
\section{Introduction}

Rare earth based materials are well known for their applications as strong permanent magnets such as  the intermetallics Nd$_2$Fe$_{14}$B and SmCo$_5$. The large orbital momentum of the rare earth $4f$ electrons is essential for the strong magnetic anisotropy of these hard magnets.
The $4f$ valence electrons are localized and they may retain spin and orbital degrees of freedom down to very low temperature, which can also result in novel ordered states such as the Kondo effect, heavy fermions, complex (sine, spiral, triple-k) magnetism, and orbital order \cite{paradigm, multipoleorders}. 
$4f$ orbitals are described to first order by their quadrupole moment although higher order moments may also order, and this can occur independently from the magnetic order.

Many functional properties such as multiferroicity, magnetoresistivity and superconductivity arise from competition between distinct order parameters. The frustration of normally dominant interactions can make higher order terms relevant. For example, ordering of the electromagnetic anapole has been observed recently in CuO \cite{copperoxide} and may be relevant to both its multiferroic properties and the superconductivity of related compounds.
Ordered states of higher rank multipole moments appear in selected compounds, such as octupoles in Ce$_{1-x}$La$_{x}$B$_6$ \cite{celab, cerium}, hexadecapoles (or triakontadipoles) in NpO$_2$ \cite{neptunium, neptheory}, and hexadecapoles in URu$_2$Si$_2$ \cite{urusitheory1, urusitheory2}. 
We have investigated the interplay between the magnetic and quadrupolar order of the $4f$ electrons in HoB$_2$C$_2$ using resonant x-ray diffraction (RXD).

RXD, with its element specificity and enhancement of the scattered intensity by up to three orders of magnitude, has been shown to be an invaluable tool in the investigation of electronic properties of materials, especially in cases where there are multiple contributions to the scattered intensity such as spin, charge, and orbital orders \cite{electronicproperties}.
In materials where the electrons are largely localized, it is advantageous to employ a local approximation such as an expansion in spherical multipole moments $Y_M^L$, and their spherical tensor representation $T_q^x$ (where $x \leq L$ and $-x \leq q \leq x$). Such spherical multipoles are parity even and have the definite time signature $(-1)^x$. We refer to time-odd multipoles as magnetic and time-even multipoles as orbital. X-ray diffraction has a convenient description in terms of spherical tensors, which includes information on the scattering geometry and polarization, and the scattering factor becomes a simple tensor product sensitive to properties of the x-ray beam and the different spherical components of the electron density at the rare earth site.

RXD is particularly successful in the investigations of exotic orders, and has for example demonstrated the existence of ordered multipole moments of up to rank 6 (the maximum allowable for a rare earth atom) in DyB$_2$C$_2$ \cite{anna, me}. While RXD studies of rare earth elements typically probe the 2p-5d dipole (E1) transition and the often much weaker 2p-4f quadrupole (E2) transition, soft resonant x-ray diffraction (SRXD) probes the 3d-4f dipole transition, and the strong interaction between the $4f$ valence and $3d$ core shells can give rise to complex spectra.

\begin{figure}
\begin{centering}
\includegraphics[width=\textwidth]{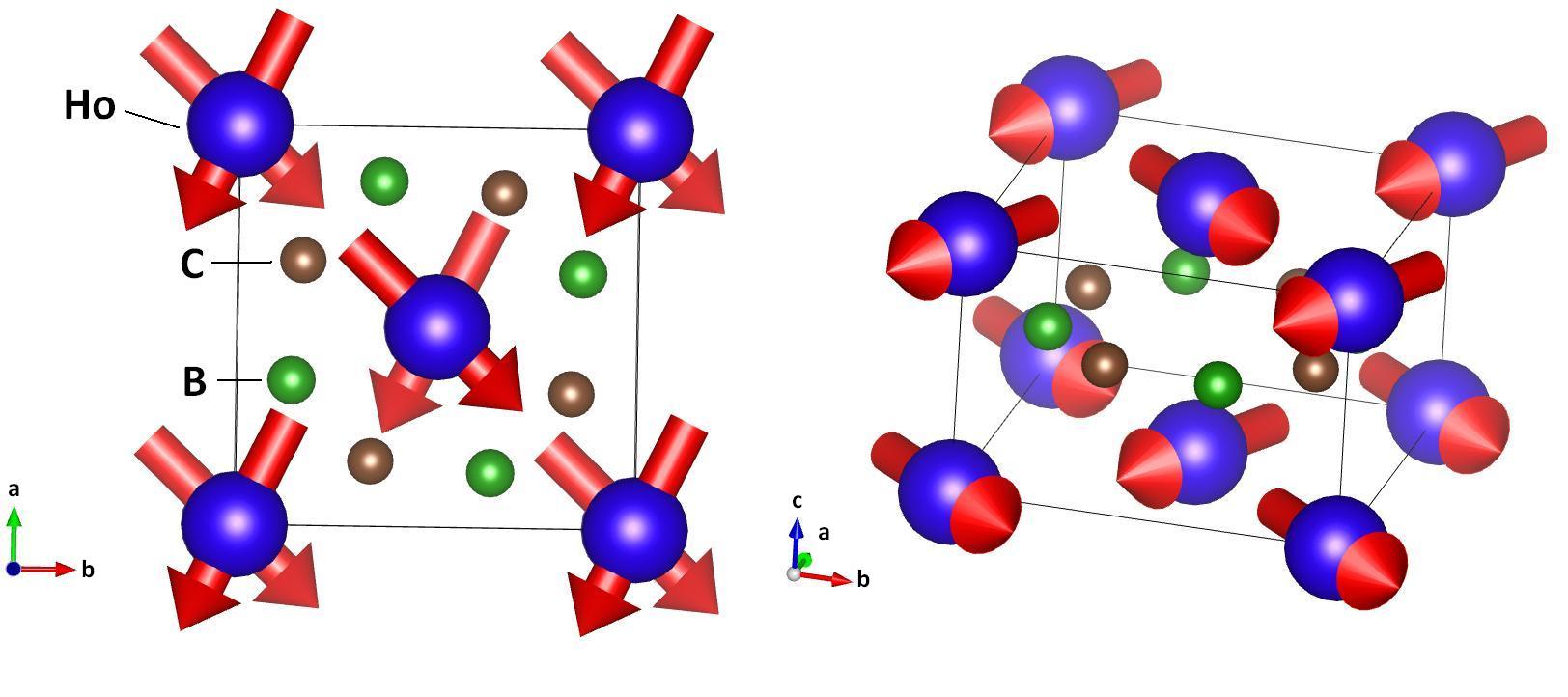}
\caption{\label{crysmagstruc} Magnetic and crystalline structure of HoB$_2$C$_2$ according to Ref. \cite{homagstruc}. Arrows indicate the direction of the Ho $4f$ magnetic moments in the low temperature Phase III. }
\end{centering}
\end{figure}

At room temperature, HoB$_2$C$_2$ crystallizes in the tetragonal space group P4$/{mbm}$ (see \Fref{crysmagstruc}) and undergoes an  antiferromagnetic (AFM) phase transition at T$_{\rm{N}} =$ 5.6 K to  the so-called Phase IV, and a further transition to a combined AFM and antiferroquadrupolar (AFQ) ordered phase at T$_{\rm{Q}} =$ 5 K (termed phase III) which has been confirmed directly with RXD at the Ho L edges \cite{matsumura} and also indirectly by ultrasonic measurement of the elastic constants \cite{ultrasound}. Below T$_{\rm{N}}$ the magnetic structure is incommensurate with ordering vector $[1+\delta, \delta, \delta']$ with $\delta = 0.11$ and $\delta'=0.04$. The magnetic structure undergoes a reorientation at T$_{\rm{Q}}$ to an AFM order having propagation vectors $k_1=[100]$ and $k_2=[01\frac{1}{2}]$ together with a ferromagnetic component $k_3=[000]$ and a weak component with $k_4=[00\frac{1}{2}]$ \cite{dyb2c2magnetism}.

This alternating arrangement of tilted moments is postulated to arise from competition between the magnetic exchange and coulomb type orbital interaction \cite{homagstruc}. It is  remarkable that the quadrupole transition occurs at a temperature lower than T$_{\rm{N}}$, as typically the onset of magnetic order removes the degeneracy of any low-lying crystal field levels. Quadrupolar transitions are most often associated with the splitting of a quartet (or pseudo-quartet) into two doublets, which is ruled out if all low-lying levels are separated into singlets by magnetic order. The reverse situation, where the doublets formed by quadrupolar ordering are again split into singlets by magnetic order, has been observed in several compounds including DyB$_2$C$_2$. Based on the release of entropy equal to $Rln3$ at the successive transitions it has been suggested \cite{Yamauchi} that the Ho ground state is a quasi-triplet consisting of a singlet and a low lying doublet, and that the doublet splits at T$_{\rm{N}}$. In this scenario, below T$_{\rm{N}}$ one of the levels of the doublet approaches the singlet forming a quasi-doublet which then gives rise to the quadrupolar transition at T$_{\rm{Q}}$. Although many investigations of this material cite an influence from quadrupole order or quadrupolar fluctuations \cite{hoinelastic}, measurement of lattice constants has inferred the presence of an octupolar order parameter \cite{ultrasound}, which together with the observation of high order multipoles in DyB$_2$C$_2$ suggests that such quantities could be observed in the low temperature Phase III.

In the present work, we performed an SRXD study of HoB$_2$C$_2$ in phase III (combined AFM + AFQ) and collected neutron and non-resonant x-ray temperature dependences for reflections relevant to the orbital ordering.  The core hole splitting formalism previously applied to the AFQ and AFM + AFQ phases of DyB$_2$C$_2$ \cite{anna, me} is employed to interpret the energy and azimuthal dependence.  In sections 2 and 3 we give an overview of the experimental details and results, followed in section 4 by a theoretical description of the resonant scattering amplitude in terms of the state multipoles of the 4f shell. In Section 5 we discuss the results for the magnitude and sign of the quadrupolar and magnetic intra-atomic interactions, and the estimations of the strength of the high order multipole moments. We also discuss anomalies in the temperature dependence that suggest the magnetic and quadrupolar order is distinct below $T_Q$. Section 6 concludes our paper with a summary of findings.

\begin{figure}
\begin{centering}
\includegraphics[width=0.75\textwidth]{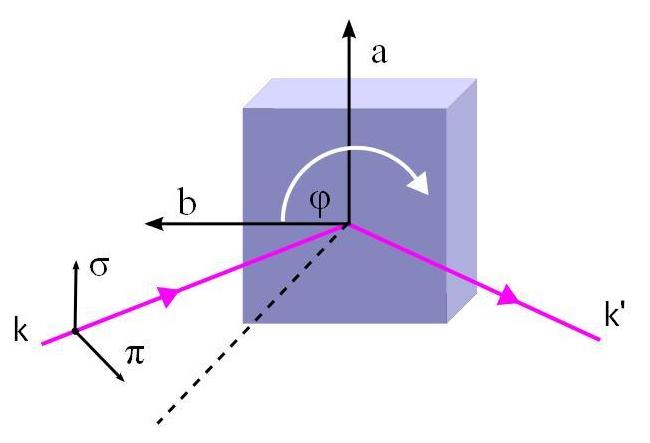}
\caption{\label{geometry} Scattering geometry used in the SRXD experiment indicating the polarization of the incident radiation, sample orientation and momentum transfer {\bf q} = {\bf k} - {\bf k'}.}
\end{centering}
\end{figure}

\section{Experimental Details}
A  HoB$_2$C$_2$ single crystal was grown by the flux-pulling method using an arc-furnace with four electrodes and cut and polished with $\left( 0 0 1 \right)$ perpendicular to the sample surface. The orbital ordering $\left( 0 0 \frac{1}{2} \right)$ reflection was recorded with $\pi$ and $\sigma$ incident polarization at the Ho M$_{4,5}$ edges of HoB$_2$C$_2$ at the UE46-PGM-1 beam-line at BESSY II synchrotron in Berlin, Germany \cite{bessybeamline}. \Fref{geometry} shows the geometry of the experiment, and a zero degree azimuthal angle corresponds to alignment of the crystallographic b-axis in the scattering plane. The sample was cooled to T = 3.5 K in a liquid helium flow cryostat and a thin aluminum thermal shield was used to maintain the temperature. Beam heating of the sample was observed which was minimized by reducing the beam intensity to a point where additional reduction did not change the measured ordering temperature. The energy dependence of the absorption was determined from the FWHM of $\theta-2\theta$ scans performed at each energy. The intensity of the $\left( 0 0 \frac{1}{2} \right)$ reflection was recorded as a function of both incident polarization and energy. At selected energies the temperature and azimuthal dependence were measured. The sample size was larger than the dimensions of the beam, making it difficult to compare absolute intensities as a function of azimuth. Consequently it is the ratio $I_\sigma / I_\pi$ that is analyzed in this work.

In addition, the temperature dependence of the $(0 0 \frac{1}{2})$ reflection was measured with neutron diffraction (ND) on the WOMBAT high intensity neutron powder diffractometer at the OPAL reactor in Sydney, Australia using 2.995 \AA~ neutrons \cite{wombat}. Further, the temperature dependence of the $(0 1 \frac{11}{2})$ reflection was measured with non-resonant x-ray diffraction (XRD) at SPring-8 on the BL19LXU beamline \cite{springbeamline}. Measurements were taken using 30 KeV x-rays and a Ge solid-state detector to eliminate higher-order harmonics.

\begin{figure}
\begin{centering}
\includegraphics[width=0.75\textwidth]{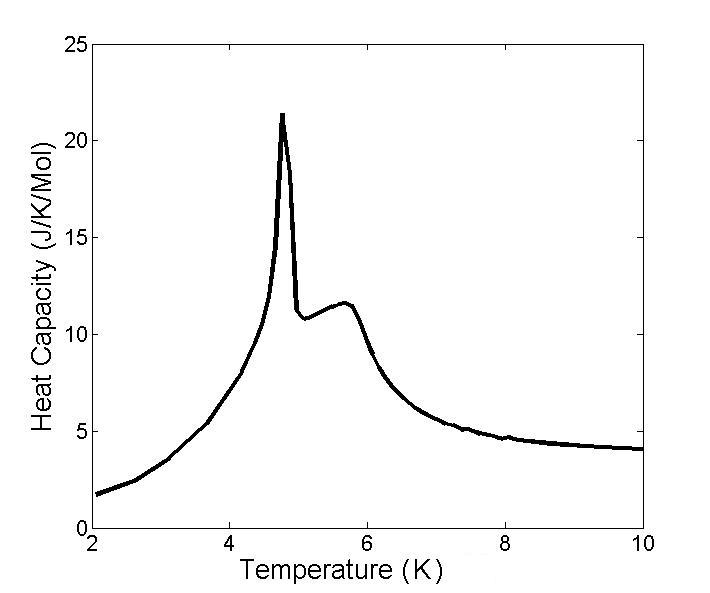}
\caption{\label{HeatCapacity} Heat capacity measurements on HoB$_2$C$_2$, showing the onset of Phase III at T$_Q$ = 4.8~K (sharp peak) and the onset of Phase IV at T$_N$ = 5.5~K (broad peak)}
\end{centering}
\end{figure}

Heat capacity was measured using a Quantum Design MPMS at Toyama University, the results of which are presented in \Fref{HeatCapacity}, showing the onset of Phase III at T$_Q$=4.8~K (sharp peak) and the onset of Phase IV at T$_N$=5.5~K (broad peak).

\section{Experimental Results}
\begin{figure}
\begin{centering}
\includegraphics[width=\textwidth]{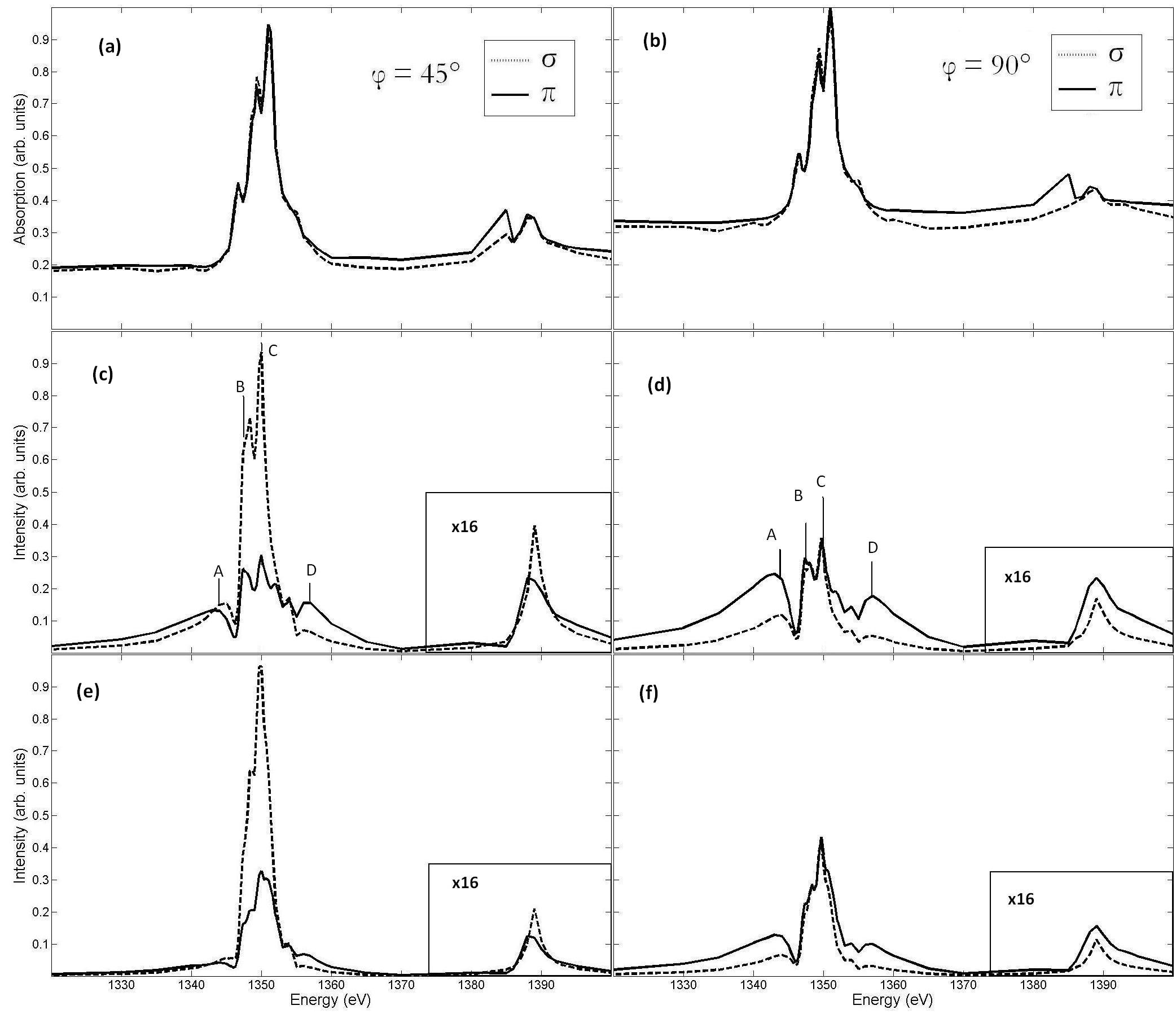}
\caption{\label{rawdata} (a) and (b), absorption spectra of the $(00\frac{1}{2})$ reflection in HoB$_2$C$_2$ determined from $\theta-2\theta$ scans with incident $\sigma$ (vertical) and $\pi$ (horizontal) polarization at 3.5 K with an azimuthal angle of 45$^\circ$ and 90$^\circ$ respectively. (c) and (d), integrated intensity of same. (e) and (f) are intensities of (c) and (d) corrected for absorption. The marked positions A-D indicate the energies at which temperature and azimuthal dependence measurements were made. Note: All data are normalized to the central peak at the M$_5$ edge in the $\sigma$ channel, and the ratio between $\sigma$ and $\pi$ channels is explicitly maintained. The intensity at the M$_4$ edge is multiplied with a factor of 16 for clarity.}
\end{centering}

\end{figure}

Figure~\ref{rawdata} shows the recorded absorption (a,b) and integrated intensity (c,d) of the space-group forbidden $\left( 0 0 \frac{1}{2} \right)$ reflection of HoB$_2$C$_2$ in phase III at an azimuthal angles of $\varphi =$45$^\circ$ and $\varphi =$90$^\circ$ as recorded with SXRD. Also shown is the absorption corrected intensity  (e,f) obtained by taking the product of the integrated intensity and the absorption. 
We draw particular attention to the qualitative difference between the $\sigma$ and $\pi$ spectra. The intensity of
the central features labeled B and C at the M$_5$ edge observed  for $\varphi = 45^\circ$ with $\sigma$ incident radiation
are much larger when compared to the other spectra, while the features A and D are largest for  $\varphi = 90^\circ$ with $\pi$ incident radiation.
This is indicative of the coexistence of two order parameters, namely the magnetic and orbital order, giving rise to interference. The significant extra intensity at $\varphi = 45^\circ$ can be attributed directly to the $A_2^2$ orbital component of the scattering as will be demonstrated later. See also Appendix C. 

\begin{figure}
\begin{centering}
\includegraphics[width=\textwidth]{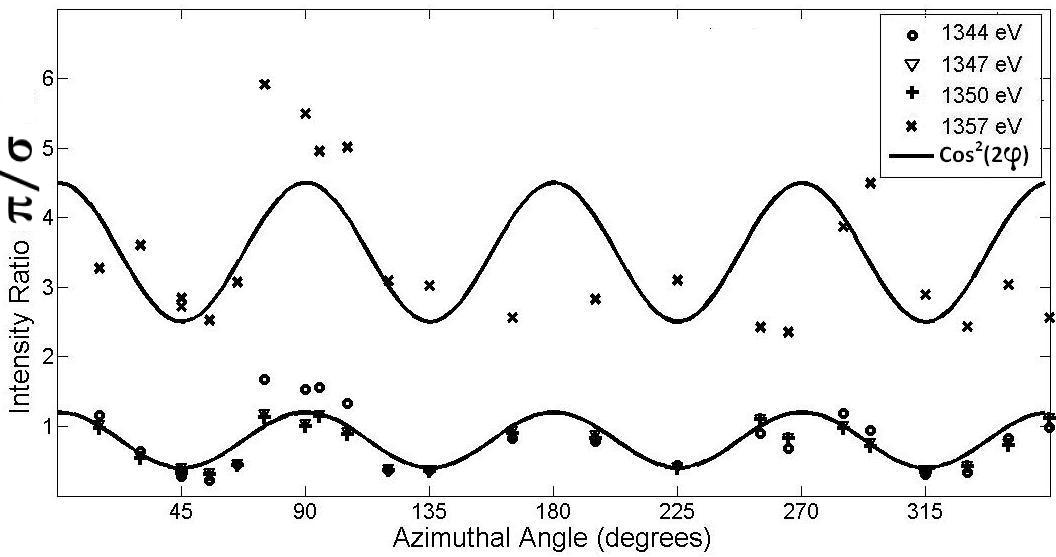}
\caption{\label{azidata} Azimuthal angle dependence of the intensity ratio $I_\sigma$ / $I_\pi$ at selected incident x-ray energies corresponding to features (A through D, respectively) in energy dependence (\Fref{rawdata}). Solid lines are proportional to $\cos^2{2\varphi}$ + constant, the expected azimuthal dependence based on the site symmetry.}
\end{centering}
\end{figure}

\Fref{azidata} shows the intensity ratio $I_\sigma / I_\pi$ as a function of the azimuthal angle $\varphi$ compared to a single order parameter possessing 2-fold rotational symmetry (i.e. the 2/m Ho site symmetry). Such a ${\cos^2{2\varphi}}$ dependence has been observed for the quadrupole E2 resonance ($2p$ to $4f$ transition) at the $L$ edge \cite{matsumura}. The azimuthal dependence at E= 1357~eV (corresponding to D in Figure 3) departs from this simple relationship, further confirming the interference of magnetic and orbital order parameters. 

\begin{figure}
\begin{centering}
\includegraphics[width=\textwidth]{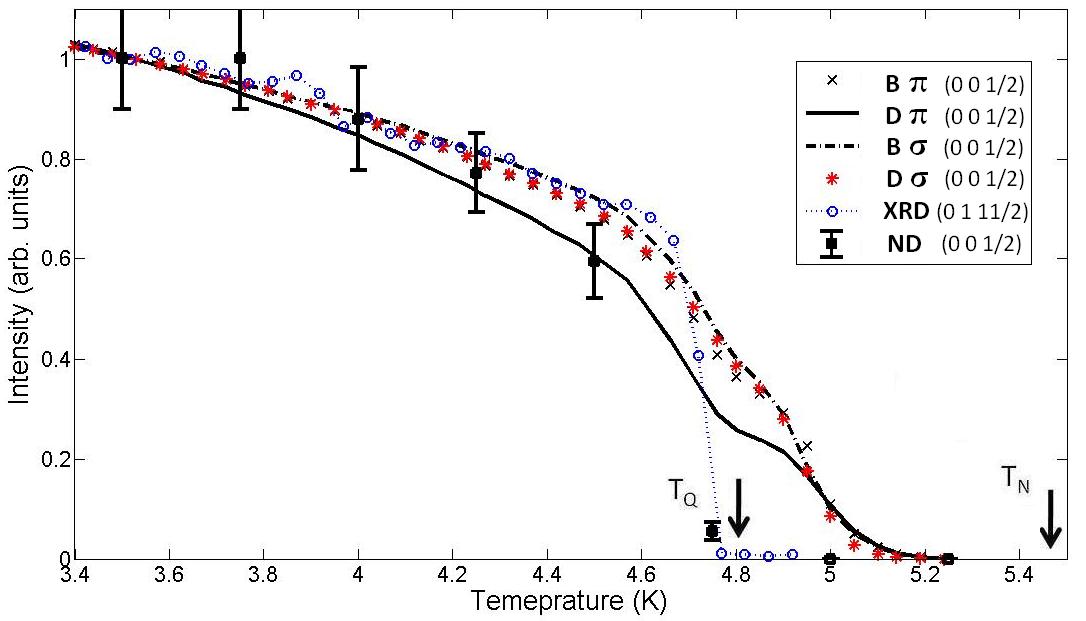}
\caption{\label{Tempdepcomparison} Measured temperature dependence of the $(00\frac{1}{2})$ reflection recorded with SRXD at selected energies corresponding to specific features B and D (Figure \ref{rawdata}), for both $\sigma$ and $\pi$ incident polarization, normalized at T =  3.5 K. The data is compared to the intensity recorded with XRD and ND.}
\end{centering}
\end{figure}

\Fref{Tempdepcomparison} compares the temperature dependence of SRXD recorded at E $\approx$ 1347 eV and E $\approx$ 1357 eV (labeled features B and D respectively in \Fref{rawdata}) for $\pi$ and $\sigma$ incident radiation and $\varphi$ = 45 $^\circ$ and 90 $^\circ$ with the recorded XRD intensity of the (0 1 11/2) reflection and ND intensity of the $(0 0 \frac{1}{2})$ reflection. 
The temperature dependences measured with the three techniques are markedly different; while SRXD has a finite intensity above T$_Q$ both $(0 0 \frac{1}{2})$ ND and $(01\frac{1}{2})$ XRD have zero intensity in this region.  The temperature dependence measured with SRXD depends on polarization, azimuthal angle, and energy, and falls between the XRD and the ND temperature dependences.  This will be discussed in sect. \ref{disc}.


\section{Theoretical Framework}
\label{theory}
It has been demonstrated \cite{anna, me} that the unique shape of the energy dependence of the $\left( 0 0 \frac{1}{2} \right)$ reflection in DyB$_2$C$_2$ can be attributed to the splitting of the 3d core states. This splitting results in multiple interfering resonators and consequently adds structure to the energy profile. The splitting of the core states is caused by a combination of the intra-atomic $4f-3d$ quadrupole interaction which splits the different $|\bar{M}|$ levels, and the intra-atomic magnetic interaction which lifts the remaining degeneracy in $\pm|\bar{M}|$, where the bar refers to the quantum numbers that label the core-hole. The relative oscillator amplitudes are determined by the 4f wavefunction, which is characterized in terms of state multipoles $\langle T_q^x \rangle$.

The X-ray scattering amplitude, the square of which is proportional to the X-ray intensity, is formed from the product of two tensors \cite{electronicproperties}
\begin{eqnarray}
\fl f = \sum_{K,Q}(-1)^{Q} X^{K}_{-Q}(2K+1)^{1/2} \sum_{q} D^{K}_{Qq}(\alpha,\beta,\gamma)\sum_{J,M}\frac{r_{\bar{J}} A_{q}^{K}(\bar{J},\bar{M})}{E-\Delta_{\bar{J}}-\epsilon^q(\bar{J},\bar{M})+i\Gamma_{\bar{J},|\bar{M}|}^{q}}
\end{eqnarray}
with ${-K}\ \leq\ {Q}\ \leq {K}$. $X_{-Q}^{K}$ is constructed from the polarization vectors of the incident and scattered X-ray beams. The Wigner D tensor $D_{Qq}^{K}(\alpha,\beta,\gamma)$ rotates $A_{q}^{K}$ (which refers to local coordinate axes of the sample), onto the coordinates of the experimental reference frame used for $X_{-Q}^{K}$ with Euler angles $\alpha,\beta,$ and $\gamma$. Our particular choice of principle axes gives $D_{Qq}^{K}(\alpha,\beta,\gamma)= D_{Qq}^{K}(0,\frac{\pi}{2},\pi-\varphi)= \e^{iq\varphi}D_{Qq}^{K}(\frac{\pi}{2})$ \cite{dydichrosim}. The Ho site symmetry restricts $q$ to $\pm 1, \pm 2$. $K=2$ corresponds to orbital order, $K=1$ to magnetic order, and $K=0$ to charge order, of which the latter is absent at the $(0 0 \frac{1}{2})$ Bragg reflection. 

The sum over the $A_{q}^{K}$ describes the series of oscillators created by the splitting of the core state \cite{anna, me}. The amplitude $A_q^K(\bar{J},\bar{M})$ of each oscillator is labeled by the total angular momentum of the core hole, $\bar{J}=\frac{3}{2}$ for the M$_4$ and $\bar{J}=\frac{5}{2}$ for the M$_5$ edge, and by the magnetic quantum number $-\bar{J} \leq \bar{M} \leq \bar{J}$. Regarding the denominator, $E$ is the photon energy, $\Delta_{\bar{J}}$ is the difference in energy between the degenerate 3d$_{\bar{J}}$ shell and the empty 4f states and $\hbar\Gamma_{\bar{J},\bar{M}}^{q}$ is the lifetime of the intermediate state. $\epsilon^q(\bar{J},\bar{M})$ is the energy shift of the core levels due to the intra-atomic quadrupole interaction $Q_{\bar{J}}$ and the magnitude of the intra-atomic magnetic interaction due to the unpaired $4f$ electrons $H_{{\rm{Int}}}$.

\begin{eqnarray}
\epsilon^q(\bar{J},\bar{M}) = {[3\bar{M}^{2}-\bar{J}(\bar{J}+1)]}Q_{\bar{J}}+g_{\bar{J}}\bar{M}\cdot H_{\rm{Int}}
\end{eqnarray}

There are six resonant oscillators at the $M_5$ edge ($\bar{M}= \pm \frac{5}{2},\pm \frac{3}{2},\pm \frac{1}{2}$) and four at the $M_4$ edge ($\bar{M}= \pm \frac{3}{2},\pm \frac{1}{2}$). The oscillator amplitudes interfere and the branching ratio between the two edges is given by the purely real mixing parameter $r_{\bar{J}}$. $A_q^K(\bar{J},\bar{M})$ is constructed from the structure factor of the chemical unit cell $\Psi_{q}^{x}$ and these two quantities are discussed in Appendicies A and B respectively. The Ho site symmetry dictates that $ A^2_2 = -A_{-2}^2$, and $ A_1^K = A_{-1}^K$, which follows from the properties of $\Psi_{q}^{x}$.  $A_1^1(\bar{J},\bar{M})$ and $A_1^2(\bar{J},\bar{M})$ are proportional to the dipole moment $\langle T_{1}^{1} \rangle$, octupole moment $\langle T_{1}^{3} \rangle$, and the triakontadipole moment $\langle T_{1}^{5} \rangle$ while $A_2^2(\bar{J},\bar{M})$ is proportional to the quadrupole moment $\langle T_{2}^{2} \rangle$, the hexadecapole moment $\langle T_{2}^{4} \rangle$, and the hexacontatetrapole moment $\langle T_{2}^{6} \rangle$. The contriutions to (1) as a function of azimuthal angle and incident polarization are discussed in Appendix C. For further theoretical details, see \cite{me}. 

\section{Discussion}
\label{disc}
\begin{figure}
\begin{centering}
\includegraphics[width=\textwidth]{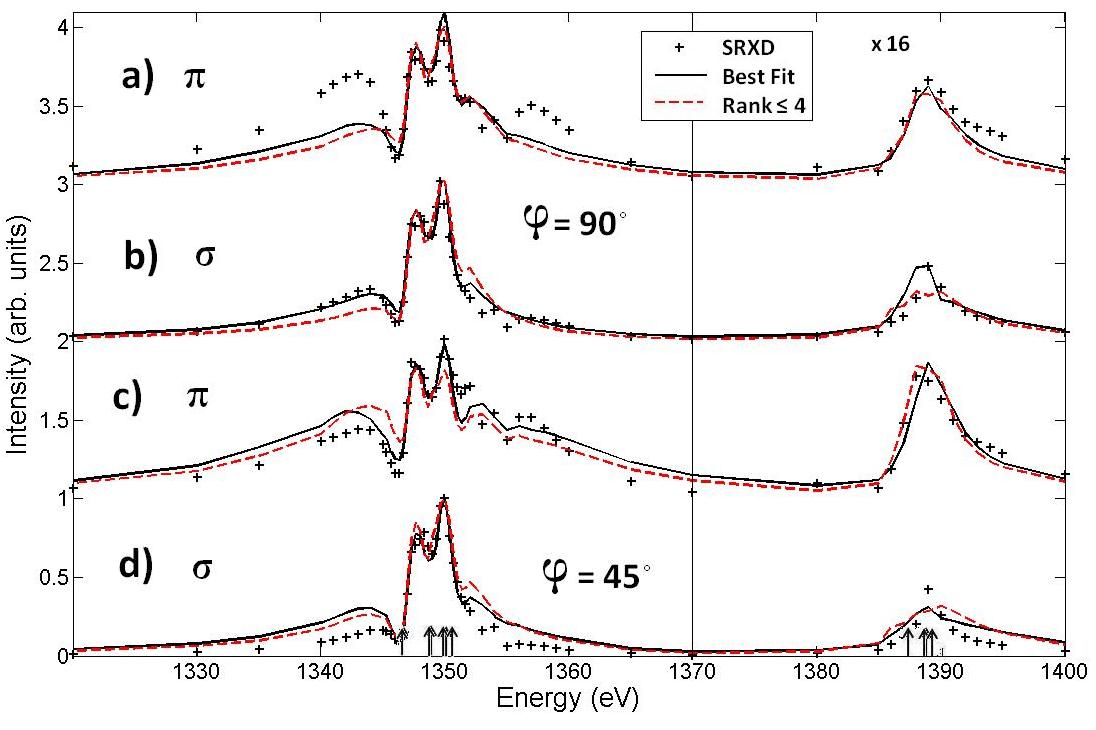}
\caption{\label{energydependancefit} Calculated energy dependence compared to normalized SRXD intensity at azimuthal angles of 90$^\circ$ (a and b) and 45$^\circ$ (c and d) for the AFM+AFQ phase at 3.5K. Presented are spectra measured with both $\pi$  (a and c) and $\sigma$ (b and d) incident radiation, shifted upwards for clarity} 
\end{centering}
\end{figure}

\begin{figure}
\begin{centering}
\includegraphics[width=0.75\textwidth]{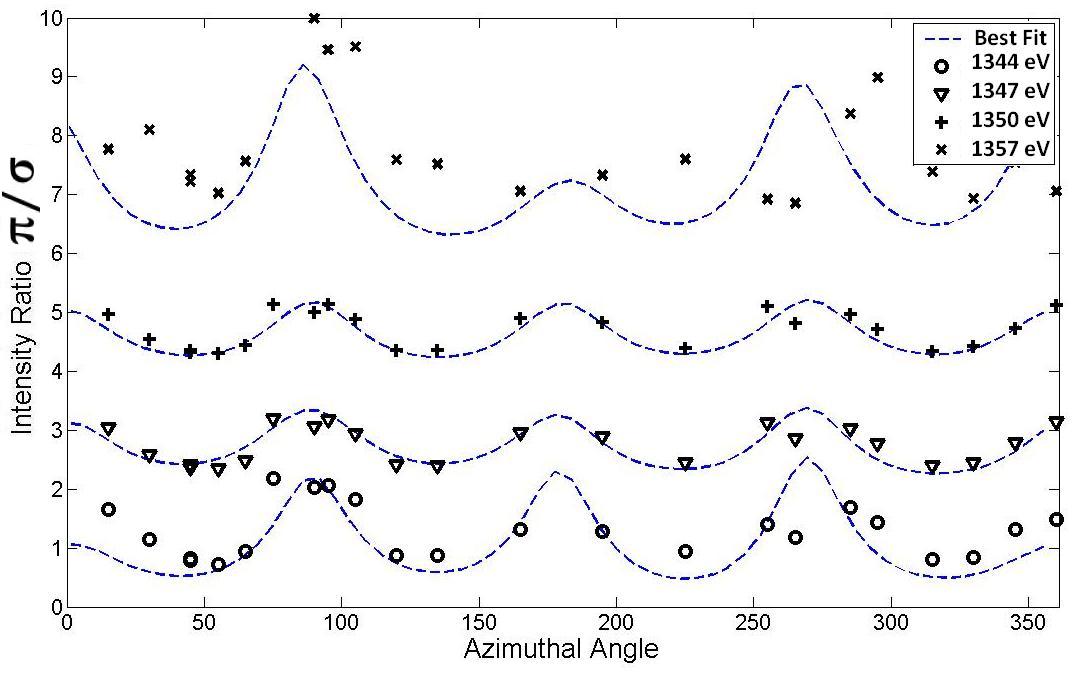}
\caption{\label{azimuthaldependencefit} Fitted azimuthal dependence of the intensity ratio I$_{\sigma}$/I$_{\pi}$ at energies corresponding to features A, B, C and D in \Fref{rawdata}, shifted upwards for clarity.}
\end{centering}
\end{figure}

\begin{figure}
\begin{minipage}[b]{0.4\linewidth}
\centering
\begin{threeparttable}[c]
\begin{tabular}{cc}
\\
Parameter & T = 3.5K (AFM + AFQ)\\
\hline
$\Delta_{5/2}$ &  ${1349~eV}$\\
$\Delta_{3/2}$ & ${1389~eV}$\\
$\Psi_{1}^{1}$  &  $0.03$\\
$\Psi_{2}^{2}$  & $0.36$ \\
$\Psi_{1}^{3}$ & $0.07$\\
$\Psi_{2}^{4}$  &  $1$\tnote{a}\\
$\Psi_{1}^{5}$  & $-0.12$\\
$\Psi_{2}^{6}$  &  $0.42$\\
Q$_{\bar{J}}$ &  ${-0.09~eV}$ \\
$|H_{Int}|$  & ${0.39~eV}$\\
r$_J$  & ${0.12}$ \\
$\Gamma_{\frac{3}{2},|\frac{3}{2}}^{2}|$ & ${1.5~eV}$ \\
$\Gamma_{\frac{3}{2},|\frac{1}{2}}^{2}|$ & ${6.3~eV}$ \\
$\Gamma_{\frac{5}{2},|\frac{5}{2}}^{2}|$ & ${0.73~eV}$ \\
$\Gamma_{\frac{5}{2},|\frac{3}{2}}^{2}|$ & ${5.7~eV}$ \\
$\Gamma_{\frac{5}{2},|\frac{1}{2}}^{2}|$ & ${1.2~eV}$ \\
$\Gamma_{\frac{3}{2},|\frac{3}{2}}^{1}|$ & ${0.40~eV}$ \\
$\Gamma_{\frac{3}{2},|\frac{1}{2}}^{1}|$ & ${1.7~eV}$ \\
$\Gamma_{\frac{5}{2},|\frac{5}{2}}^{1}|$ & ${1.6~eV}$ \\
$\Gamma_{\frac{5}{2},|\frac{3}{2}}^{1}|$ & ${12.4~eV}$ \\
$\Gamma_{\frac{5}{2},|\frac{1}{2}}^{1}|$ & ${2.7~eV}$ \\
\hline
\end{tabular}
\begin{tablenotes}
\item [a] This value is set to 1
\end{tablenotes}
\end{threeparttable}
\end{minipage}
\hspace{0.5cm}
\begin{minipage}[b]{0.6\linewidth}
\centering
\includegraphics[width=0.75\textwidth]{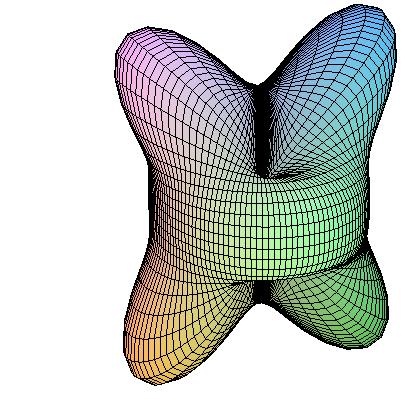}
\caption{\label{derivedorbitals} (left) Fitted Parameters in the AFM + AFQ phases of HoB$_2$C$_2$. (right) Reconstruction of the angular dependence of the $4f$ charge density described by the values of $ \Psi_q^x $ in the table. }
\end{minipage}
\end{figure}

The theory detailed in section \ref{theory}  is used to fit simultaneously the measured energy dependence for $\sigma$ and $\pi$ incident polarization and the measured azimuthal angle dependences via global optimization using a simple genetic algorithm method and a pattern search optimization of the genetic algorithm result. Although such a method cannot be guaranteed to locate the global minimum, by repeating the fitting process until the best solution has been obtained multiple times we obtain a statistical uncertainty ($2\sigma$) of better than 10\% for all parameters except the linewidth  $\Gamma_{\frac{3}{2},|\frac{3}{2}|}^1$ which has an uncertainty of 20\%. The best fit is shown in  \Fref{energydependancefit} and  \Fref{azimuthaldependencefit} for the energy and azimuthal angle dependence, respectively.  The absorption correction is added to the calculated spectra so it can be compared with the as-measured data, and the oscillator positions are indicated in  \Fref{energydependancefit} with arrows. The parameters obtained in the fit, and the associated angular distribution of the $4f$ electronic charge density, are shown in \Fref{derivedorbitals}. The branching ratio r$_{3/2} = 0.12$ calculated for the free ion \cite{vanderlaan} gives an excellent fit. We find that the intra-atomic quadrupole interaction Q$_{5/2}$ = $-0.09~eV$, and the intra-atomic magnetic interaction $|H_{\rm{Int}}|$ = $0.39~eV$. Both of these values are largely independent of the choice of wavefunctions and linewidths, and the order of magnitude of H$_{\rm{Int}}$ is in line with that expected from atomic calculations of 3$d$-4$f$ exchange integrals \cite{vanderlaan}. It is important to note that whilst oscilators with $\pm M$ will be shifted in opposite directions by $H_{\rm{Int}}$, there is no way to distinguish between them and so only the magnitude of $H_{\rm{Int}}$ can be used as a fitting parameter, its sign being arbitrary. 

Variations in $\langle T_{q}^{x} \rangle$ can be partially compensated by the line widths to yield a similar result. The linewidth for the oscillators $\Gamma_{\bar{J},|\bar{M}|}^q$ vary from $\approx 0.5~eV$ to $12~eV$ and vary substantially between time even and time odd contributions. The hexadecapole $\Psi_{2}^{4}$ is the dominant contribution and is normalized to 1, giving $\Psi_{1}^{1} = 0.03$, $\Psi_{2}^{2} = 0.36$, $\Psi_{1}^{3} = 0.07$, and $\Psi_{1}^{5} = -0.12$, $\Psi_{2}^{6} = 0.42$. It is notable that the lower rank magnetic contributions (the octupole and dipole) are small relative to the time-even contributions. Whilst the actual ordered moment is given by $\langle T_{q}^{x}\rangle$, the magnetic contributions (odd q) to the scattering are sensitive to $ \langle T_{q}^{x}\rangle' -  \langle T_{q}^{x}\rangle''$ (where $\langle \ldots \rangle'$ and $\langle \ldots \rangle''$ represent respectively the real and imaginary parts of the atomic tensors, see Appendix A) which can potentially be small when the ordered moment is large, in the case where the real and imaginary parts are of similar size. This coincidence cannot be ruled out, as neutron diffraction studies have shown that the magnetic moment is large, although a substantial portion of the ordered moment propagates along directions other than the $(0 0 \frac{1}{2})$ \cite{rbcmagnetism}. Additionally, the moment $\langle J \rangle$ measured with neutron diffraction is not precisely the same quantity as, for example, the $\langle T_{1}^{1}\rangle$ magnetic dipole tensor which is a linear combination of $\langle \bf{S} \rangle$, $\langle \bf{L} \rangle$, and $\langle 3\bf{R}(\bf{R} \cdot \bf{S})-\bf{S} \rangle$ \cite{electronicproperties}. Barring this accidental case of near-cancellation, the relatively small contribution of the ordered magnetic moments (of lower rank) can be attributed to the frustrated interactions of the system, as the magnetic and orbital exchanges strongly compete with one-another.

For comparative purposes, we also plot a best fit in \Fref{energydependancefit} without the higher order multipoles (i.e. rank 5 and 6) that are not directly observable via any other means, and retain only terms of rank $\leq 4$ that are observable via an E2 resonance. The inclusion of higher rank terms significantly improves the description of the data in the pre-edge region, particularly where the intensity dips towards zero. 

The values of Q$_{5/2}$ and $|H_{\rm{Int}}|$ obtained by this analysis are smaller in magnitude and of the same sign as those obtained from the analysis of the AFM+AFQ phase of DyB$_2$C$_2$ where they were found to be $-0.42$ and $0.98$ eV respectively \cite{me}. The interaction between the $3d$ and $4f$ shells is expected to be similar for both Ho$^{3+}$ and Dy$^{3+}$ as they contain 9 and 10 $4f$ electrons respectively, and the radial extent of the $4f$ shells are almost identical.
This suggests both the magnetic and quadrupolar moments of the Ho $4f$ shell are smaller than for the Dy $4f$ shell in $R$B$_2$C$_2$. The two materials have a similar magnetic structure with four magnetic sublattices characterized by the propagation vectors [1 0 0], [0 0 $\frac{1}{2}$], [1 0  $\frac{1}{2}$] and [0 0 0] , and are inferred to have a similar pattern of orbital ordering on this basis, although HoB$_2$C$_2$ has a much lower ordering temperature \cite{homagstruc, rbcmagnetism}. The magnetic structures are compared in \cite{rbcmagnetism} and the component of ${\bf k}$  along (0 0 $\frac{1}{2}$) is found to be smaller in HoB$_2$C$_2$ compared to DyB$_2$C$_2$. Moreover, resonant diffraction at the Ho L-edge has shown that the quadrupolar moment in HoB$_2$C$_2$ is smaller than in DyB$_2$C$_2$ \cite{matsumura}. 

The temperature dependences recorded with SXRD, ND and XRD show noticeable differences due to their different structure factors (Fig. \ref{Tempdepcomparison}). 
The structure factor of the (0 1 $\frac{11}{2}$) reflection recorded with XRD is proportional to \cite{tbanna} 
\begin{equation}
F_{0 1 \frac{11}{2}} \propto [ \cos (2 \phi_1 ) - \cos ( 2 \phi_3 ) ] \langle Q_{0 1 \frac{11}{2}} \rangle
\label{eq_xd}
\end{equation}
where $ \phi_n$ is the angle of the principal axis of the Ho orbital at site $n$ with respect to the (1 0 0) axis in the $ab$ plane and $\langle Q_{0 1 \frac{11}{2}} \rangle$ is the expectation value of the Ho $4f$ time-even multipolar moment which is proportional to a weighted sum of $\Psi_2^2$,  $\Psi_2^4$ and  $\Psi_2^6$ \cite{tbanna}. 
Thus the diffracted intensity observed with XRD is proportional to the orbital, or time-even, order of the Ho $4f$ shell.
In contrast, the intensity at (0 0 $\frac{1}{2}$) observed with ND is proportional to the magnetic dipole moment of the Ho $4f$ shell, in particular to the magnetic moment components in the $ab$ plane as those are perpendicular to the scattering vector. Figure \ref{Tempdepcomparison} shows that the temperature dependence observed with these methods is distinct and this suggests that the temperature dependence of the orbital and magnetic order parameters are themselves distinct. Both order parameters shows a power law dependence $I \propto (\frac{T_Q - T}{T_Q})^{2\beta_x}$ with the magnetic critical exponent $\beta_M \approx 0.35$, while the orbital order increases more abruptly with $\beta_Q \approx 0.2$. 
The temperature dependence of the SXRD intensity recorded at the (0 0 $\frac{1}{2}$) depends on energy, polarization and azimuthal angle and has the ND and XRD curves as lower and upper extremities, respectively (\Fref{Tempdepcomparison}). This supports the claim that SXRD measures the interference of the magnetic and orbital order parameters of which the relative contribution depends on the energy, incident polarization and azimuthal angle.

To illustrate this point, \Fref{AKQplot} presents the energy dependence of the magnetic contributions, $A_1^1$ and $A_1^2$, and orbital contribution, $A_2^2$, to the scattered intensity $\varphi = 90^\circ$ and $\pi$ incident radiation as well as $\varphi = 45^\circ$ and $\sigma$ incident radiation. While in the first the magnetic part dominates for feature D, in the latter the orbital part dominates in feature B, consistent with the temperature dependencies presented in Fig. \ref{Tempdepcomparison} when compared to neutron and x-ray diffraction respectively.

\begin{figure}
\begin{centering}
\includegraphics[width=0.75\textwidth]{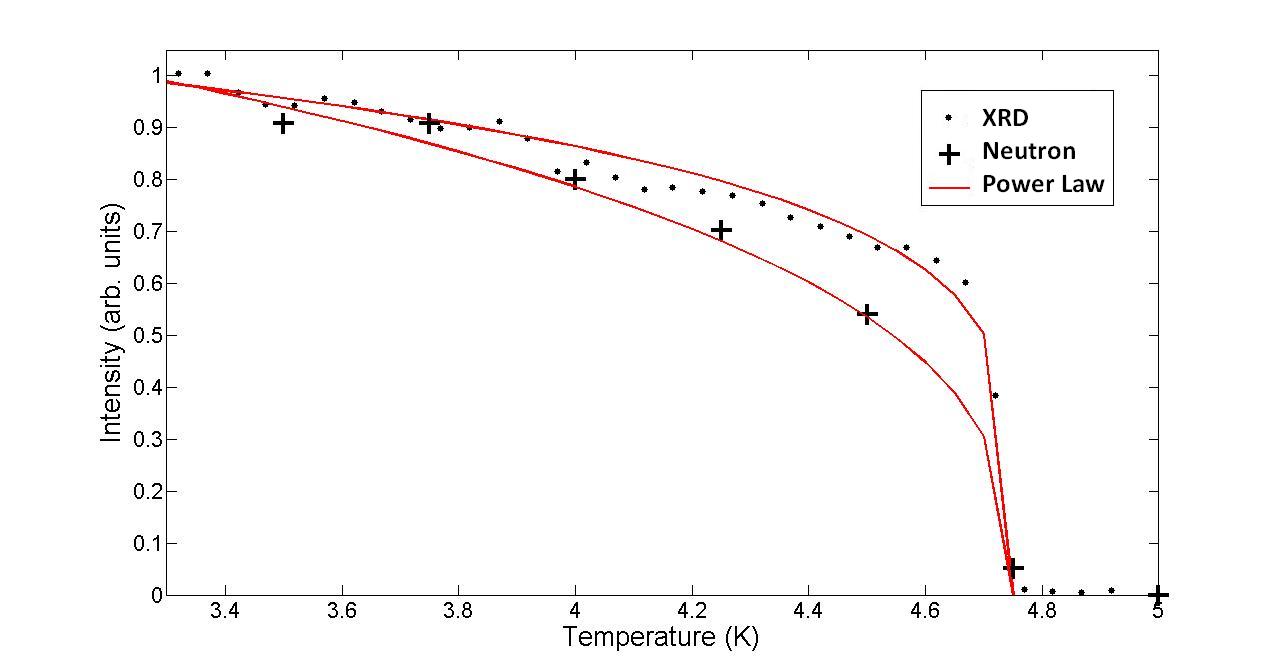}
\caption{\label{criticalexponent} temperature dependence of the neutron and x-ray intensities, normalized to 3.3 K, with illustrative power law fits $I \propto (\frac{T_Q - T}{T_Q})^{2\beta}$. The power law curves correspond to $\beta_Q \approx 0.2$ and $\beta_M \approx 0.35$ }
\end{centering}
\end{figure}

\begin{figure}
\begin{centering}
\includegraphics[width=\textwidth]{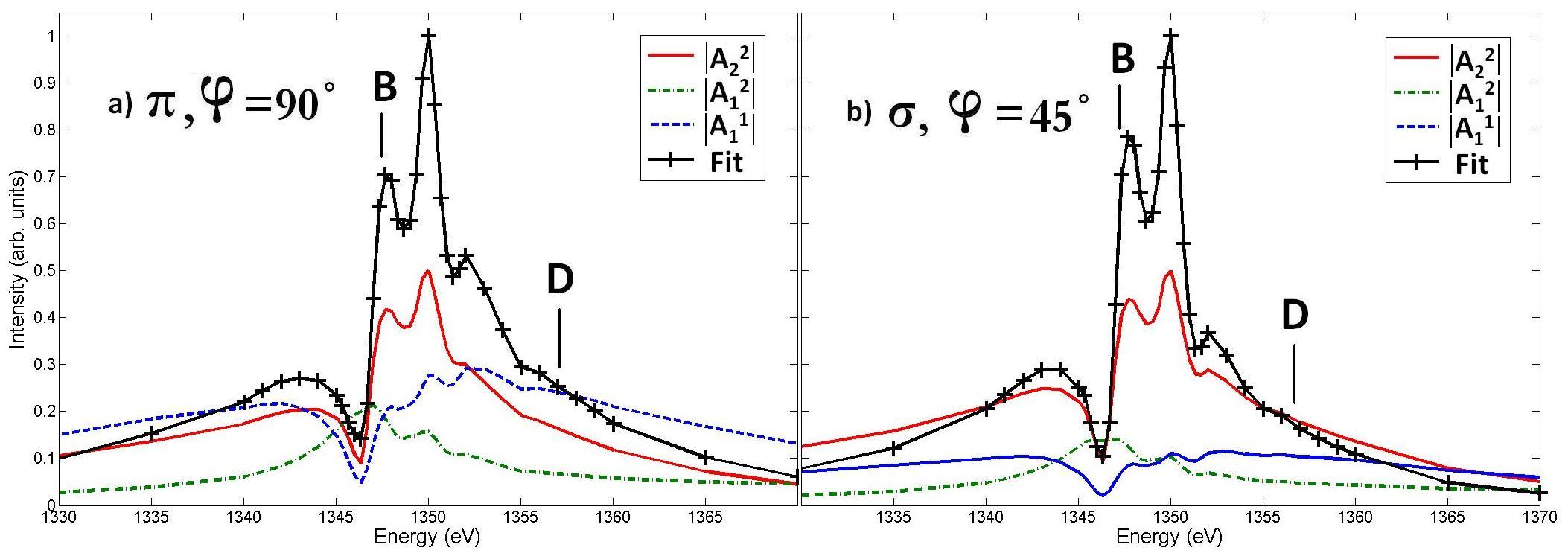}
\caption{\label{AKQplot} The energy dependence of the magnetic parts, $A_1^1$ and $A_1^2$, and orbital part, $A_2^2$, of the (0 0 $\frac{1}{2}$) reflection for  (a) $\varphi = 90^\circ$ and $\pi$ incident radiation and (b) $\varphi = 45^\circ$ and $\sigma$ incident radiation. B and D are labeled in accordance with \Fref{rawdata}. Note that the orbital contribution is dominant for feature B in (b) while the magnetic contribution is dominant for feature D in (a), agreeing with the measured temperature dependences in \Fref{Tempdepcomparison}.}
\end{centering}
\end{figure}

Distinct temperature dependences within a single resonance profile has been reported in La$_{0.5}$Sr$_{1.5}$MnO$_4$ \cite{COvsJT}. At particular energies of the Mn L$_{2,3}$-edges,  the temperature dependence of the orbital order reflection ($\frac{1}{4}$ $\frac{1}{4}$ 0) resembles that of the orbital order parameter, while at other energies the temperature dependence resembles that of the Jahn-Teller distortion. RXD on the (1 0 2) and $(1 0 \frac{3}{2})$ reflections in TbB$_2$C$_2$, performed at the Tb L-edge, have shown two distinct critical exponents, tentatively assigned to the magnetic and orbital order parameters \cite{tb_criticalexponents}. In this case, $\beta_M \approx 0.3$, and $\beta_Q \approx 0.45$, suggesting that the orbital order is induced by (and possibly competes with) the magnetic order. Distinct timescales have also been observed in the paramagnetic and paraorbital state of DyB$_2$C$_2$, where it has been observed that the $4f$ orbital moments fluctuate faster than the magnetic moments \cite{staub_prl_2005}. For HoB$_2$C$_2$ it has been suggested \cite{homagstruc} that the magnetic order induces the orbital order at T$_Q$ due to the formation of a quasi doublet ground somewhere below T$_N$ due to near-crossing of two crystal field levels in the presence of magnetic exchange. Orbital fluctuations can inhibit the AFQ order and inelastic neutron scattering performed on a HoB$_2$C$_2$ sample \cite{hoinelastic} noted a broadening of the peaks at T$_Q$ which has been attributed to strong orbital fluctuations at the critical temperature. In contrast to the case of TbB$_2$C$_2$, we find that those components of the scattered intensity whose contribution is largely orbital have the smaller critical exponent, and a consequently sharper temperature dependence. This indicates that the magnetic reorientation is at least partially induced by the orbital ordering. There are not enough data points from the neutron scattering data to attempt a proper measurement of the critical exponent, but an illustrative attempt has been made to convey the differences between the neutron and x-ray scattering data in \Fref{criticalexponent}.

Remarkably, a finite intensity has been observed with SXRD up to T = 5 K while the intensity observed with both the XRD and ND is zero (\Fref{Tempdepcomparison}). At present the reason for this is unclear. SXRD is sensitive to all directions of the magnetic moments, as well as to higher order time-odd multipole moments whereas ND measures the in plane dipole moments only. Thus a  dipole moment along the $c$-axis or higher order time-odd multipole moments $\Psi_1^3$ or $\Psi_1^5$ can give rise to finite SXRD while ND and XRD remain zero. This implies a magnetic reorientation transition from the incommensurate long range magnetic order between T$_N$ and T = 5 K to commensurate magnetic order below T = 5 K to combined magnetic and quadrupolar order below T$_Q$. However, the energy dependence observed with SXRD at T = 4.9 K (not shown) displays the characteristics of combined orbital and magnetic order. Possibly, short range correlations associated with the transition at T$_Q$ may become pinned at the surface leading to an observable signal above T$_Q$. SXRD probes the surface region of the sample ($\approx$ 200 nm) while XRD and ND probe $\approx$ 100 $\mu$m and the whole sample respectively. An additional possibility arises from the inherent uncertainty in measuring the sample temperature, and it cannot be ruled out that the intensity measured at 5 K with SRXD should coincide with the intensity measured at 4.8 K with XRD and ND, although it is problematic to explain the presence of the jump in intensity at the lower temperature if this is the case. Therefore we cannot exclude that the SXRD signal observed above T$_Q$ is due to either a surface effect or the uncertainty in the absolute temperature. We emphasize that, if present, such effects cannot account for the differences observed in the temperature dependencies. Moreover, a difference in the measured ordering temperature of 0.2~K has a minor effect on the SXRD recorded at T = 3.4~K, which is well below T$_Q$.

\section{Conclusion}
HoB$_2$C$_2$ exhibits an ordered phase (phase III) with both antiferromagnetic (AFM) and antiferro-orbital (AFQ) order below T$_{\rm{Q}}$ = 4.8 K. Soft resonant x-ray Bragg diffraction experiments were performed at the Ho M$_{4,5}$ edges and the energy dependence of the space group forbidden $(00\frac{1}{2})$ reflection was measured in this low temperature AFM+AFQ phase. The energy dependence of the $(00\frac{1}{2})$ reflection is modeled successfully by including an intra-atomic core-hole interaction parameterized by quadrupole and magnetic interactions. A pseudo-multiplet structure was included, allowing for the interference between oscillators that have different properties under time-reversal due to their having different excited-state lifetimes. The Ho $4f$ multipole moments of rank 1 (dipole) to 6 (hexacontatetrapole) were determined, with the contributions from rank 1 and 3 time-odd multipoles found to be small compared to the dominant contribution of the hexadecapole (rank 4). 

Comparison of the temperature dependence of the orbital ordering reflections recorded with SXRD, XRD and ND shows that the magnetic and quadrupolar order parameters are distinct. The magnetic order exhibits a power-law dependence below T$_Q$ with a critical exponent that is smaller than that of the orbital order. We conclude the quadrupolar interaction is strong, but quadrupolar order only occurs when the magnetic order results in a near-crossing of the crystal field levels, producing a quasi doublet ground state. This is consistent with strong quadrupolar fluctuations present above T$_Q$ and the  lock-in like transition of the orbitals at T$_Q$.

\section{Acknowledgements}

The authors wish to acknowledge travel funding provided by the International Synchrotron Access Program  managed by the Australian Synchrotron and funded by the Australian Government. This work was partly supported by Grants-in-Aid for Scientific Research (A) No. 21244049 from JSPS and the non-resonant synchrotron radiation experiments were performed at beam line 19LXU in SPring-8 with the approval of RIKEN (Proposal No. 20090012). Neutron diffraction experiments were performed at the WOMBAT beam line of OPAL with the assistance of AINSE and approval ANSTO (Proposal ID P1242).

\appendix

\section{The atomic scattering amplitude $A_q^K{\bar{J},\bar{M}}$ and the reduced matrix elements $R^{K}(r,x)$}

$A_q^K(\bar{J},\bar{M})$ is constructed from the structure factor of the chemical unit cell $\Psi_{q}^{x}$ and reduced matrix elements $R^{K}(r,x)$ according to 

\begin{equation}
\fl A_q^K(\bar{J},\bar{M})=
(-1)^{\bar{J}-\bar{M}}\sum_{r}(2r+1)
\left(
\begin{array}{ccc}
\bar{J} & r & \bar{J} \\
-\bar{M} & 0 & \bar{M} \\
\end{array}
\right) 
\sum_{x} 
\left(
\begin{array}{ccc}
K & r & x \\
-q & 0 & q \\
\end{array}
\right) 
R^{K}(r,x)\Psi_{q}^{x}
\end{equation}
where the symbol in brackets is a 3jm symbol, with $r=0,1,\ldots,2\bar{J}$, $x= | K - r |,\ldots,| K + r |$ and $q+x$ and $r+x$ are both even integers. 
The reduced matrix elements $R^{K}(r,x)$ are calculated for the Ho$^{3+}$ ground state of ${^5}I_{8}$ according to the relation found in the appendix of \cite{me}. A Matlab program for calculating these reduced matrix elements for a valence configuration of n$_h$ $p$,$d$, or $f$-electron holes appropriate to either an $E1-E1$ or $E2-E2$ event is available from the corresponding authors upon request.

\section{Ho site symmetry and magnetic structure}

The structure factor is calculated using the magnetic structure presented in \Fref{crysmagstruc}. For simplicity the small c-axis canting described by \cite{rbcmagnetism} is excluded and we have the following structure factor of the atomic tensors.
\begin{eqnarray}
\Psi_{q}^{x}=(1-e^{\frac{-20i\pi q}{180}}e^{\frac{i\pi q}{2}})\{\langle T_{q}^{x}\rangle-e^{\frac{20i\pi q}{180}}e^{\frac{i\pi q}{2}}\langle T_{-q}^{x}\rangle \} \\
\Psi_{2}^{x}= -i\left(1.29\langle T_{2}^{x} \rangle' -3.53\langle T_{2}^{x} \rangle''\right)\\
\Psi_{1}^{x}= -i\left(1.88i\langle T_{1}^{x} \rangle' - 1.32\langle T_{1}^{x} \rangle''\right)
\end{eqnarray}
where $\langle \ldots \rangle'$ and $\langle \ldots \rangle''$ represent respectively the real and imaginary parts of the atomic tensors. We note that the fitting process includes $\Psi_{q}^{x}$ directly.

\section{Contributions to the scattering factor as a function of azimuthal angle}

The energy-independent component of the structure factor for contributions to the scattering is given (with the quantum numbers $\bar{J}$ and $\bar{M}$ of $A_q^K$ omitted for the sake of brevity) as
\begin{eqnarray}
\fl f_{\sigma,\pi} = -2\sqrt{5}iA_{2}^{2}\cos(2\varphi)\sin(\theta)+\sqrt{5}iA_{1}^{2}\sin(\varphi)\cos(\theta)\nonumber\\
-\sqrt{3}A_{1}^{1}\cos(\varphi)\cos(\theta)\\
\fl f_{\sigma,\sigma}= 2\sqrt{5}iA_{2}^{2}\sin(2\varphi)\\
\fl f_{\pi,\sigma} = -f_{\sigma,\pi}\\
\fl f_{{\pi,\pi}} = \sqrt{5}iA_{2}^{2}\sin(2\varphi)(1-\cos(2\theta))+\sqrt{3}A_{1}^{1}\sin(\varphi)\sin(2\theta)
\end{eqnarray}
which gives for $\varphi = 45^\circ$ and $2\theta \approx 90^\circ$
\begin{eqnarray}
\fl I_{\sigma}= \frac{1}{4}|\sqrt{5}iA_{1}^{2} - \sqrt{3}A_{1}^{1}|^2 +20 |A_{2}^{2}|^2  \\
\fl I_{\pi}=  \frac{1}{4}|\sqrt{3}A_{1}^{1} -\sqrt{5}iA_{1}^{2}|^2 +|\sqrt{5}iA_{2}^{2} - \sqrt{\frac{3}{2}}A_{1}^{1}|^2
\end{eqnarray}
and for $\varphi = 90^\circ$ with $2\theta \approx 90^\circ$
\begin{eqnarray}
\fl I_{\sigma}=  \frac{5}{2}|2iA_{2}^{2}+iA_{1}^{2}|^2 \\
\fl I_{\pi}= 3|A_{1}^{1}|^2 + \frac{5}{2}|2iA_{2}^{2}+iA_{1}^{2}|^2
\end{eqnarray}

\end{document}